\documentstyle[aps,twocolumn]{revtex}

\begin{document}
\date{\today}

\title{Current Statistics for  Quantum Transport
through Two-Dimensional Open Chaotic Billiards}
\author{Alexander I.~Saichev,$^{1,2}$ Hiromu Ishio,$^{1,3}$
Almas F.~Sadreev,$^{1,4}$ and Karl-Fredrik Berggren$^{1}$}
\address{$^{1}$Department of Physics and Measurement Technology,
Link\"oping University, S-581 83 Link\"oping, Sweden\\
$^{2}$Department of Radiophysics, Nizhny Novgorod State University,
Gagarin prosp. 23, 603600, Nizhny Novgorod, Russia\\
$^{3}$School of Mathematics, University of Bristol,
University Walk, Bristol BS8 1TW, UK\\
$^{4}$Kirensky Institute of Physics, 660036, Krasnoyarsk, Russia}
\date{\today}
\maketitle
\begin{abstract}
The probability current statistics of
two-dimensional open chaotic ballistic billiards
is studied both analytically and numerically.
Assuming that the real and imaginary parts of the scattering wave function
are both random Gaussian fields,
we find a universal distribution function for
the probability current. In by-passing we recover previous analytic
forms
for wave function statistics.
The expressions bridge the entire region from  GOE to
GUE type statistics. Our analytic expressions
are verified  numerically by explicit quantum-mechanical calculations of
transport through a Bunimovich billiard.
\newline
 PACS 05.45.Mt,05.60.Gg,73.23.Ad
\end{abstract}

\section{Introduction}
For quantum chaotic closed system
it is well known that the statistical properties of the energy levels
are described by random matrix theory (RMT)\cite{HJStockmann}.
They follow the Gaussian orthogonal ensemble (GOE) and the Gaussian unitary
ensemble (GUE),
depending on whether time-reversal symmetry (TRS) of a system is preserved
or not. In the same way the wave function statistics obeys different laws in
the  two cases.
Let the scaled local density be  $\rho({\bf r})=A|\psi({\bf r})|^2$ where
 $\psi({\bf r})$ is the normalized  wave function
and $A$ is the area (volume) of the system.
As prescribed by GOE
the probability distribution
is the well-known Porter-Thomas (P-T) distribution
$P(\rho)=\left(1/\sqrt{2\pi \rho}\right)\exp(-\rho/2)$
  when TRS is present (the Hamiltonian $H$ is invariant
under time reversal $\hat T (t\rightarrow-t)$ and $\psi$ may be chosen real).
On the other hand, the distribution takes the exponential Rayleigh form
$P(\rho)=\exp(-\rho)$ as
described by GUE when TRS is broken ($H$ is not invariant on $\hat T$
and $\psi$ must be complex). It is easy to understand qualitatively  why
the statistics is so different in the two cases, for example, why
small values of $\rho$ have a much larger weight in GOE than in GUE.
In the first case the real wave function vanishes along nodal lines in
two-dimensional (2D) systems (surfaces in 3D). On the other hand, $\psi$
is complex in the second case
and vanishes only at nodal points (lines in 3D) resulting
in less probability for small $\rho$.
 Depending on the relative weights of
the orthogonal real and imaginary parts of $\psi$ one can also define
intermediate statistics that applies to the entire crossover region
from GOE to GUE\cite{zyczkowski,kanzieper}.

We now consider what happens when the system is made open, for example
 by attaching electron leads to some exterior reservoirs and 
 a stationary current through the system is induced
 by applying suitable voltages to the reservoirs. The additional flexibility
  gained in this way leads to a number of
interesting cases for the wave function statistics. Let us first look
 at the case when there is no current flow. The statistics will then be
the  same as for the closed system above, {\it i.e.}, the kind of statistics simply
  depends on whether
  the Hamiltonian is invariant under $\hat T$ or not. On the other hand,
  if there is a stationary current via the leads we have to deal
  with a  scattering wave function. This function, which  must be complex,
  is written in 2D as
\begin{equation}
\label{psi}
\psi(x,y)=u(x,y)+iv(x,y).
\end{equation}
and satisfies $(\nabla^2+k^2)\psi=0$.
\noindent Even if the Hamiltonian itself is invariant under
$\hat T$, the statistics will not follow GOE since the 
scattering wave function is
complex. Because of
the boundary conditions associated with the scattering wave function
it is no longer an eigenstate of the usual time reversal operator $\hat C\hat T$
where
 $\hat C$ is complex conjugation of $\psi(\bf r)$.
We will show how the two Gaussian random fields $u$ and $v$ are identified
to recover the intermediate statistics discussed above
  and how  a universal distribution for the probability current density can be found.
 We  will also compare theory with explicit
  numerical calculations for  an open Bunimovich stadium.

\section{Theory}

In the following derivation of the wave function and probability current
statistics we assume that the real and imaginary parts of
$\psi$ can be viewed as two independent isotropic Gaussian fields. An explict 
example of such a state is given in ref. \cite{RPnini} in the form of a Berry-type
wave-chaotic function.
In general the assumption of independent fields can only make sense if we first extract a
common phase factor. This feature will turn out to be most useful.
Let us introduce the notations
\begin{equation}
\label{disp}
\langle  u^2\rangle  = \sigma_u^2, ~\langle  v^2\rangle  = \sigma_v^2,
~\langle  uv\rangle  = \gamma,
\end{equation}
\begin{equation}
\label{dispersion}
\sigma^2 = \sigma_u^2+\sigma_v^2=\langle  |\psi|^2\rangle ,\\
~\langle  u\rangle =0, ~\langle  v\rangle =0.
\end{equation}

\noindent We define the averages as
\begin{equation}
\label{average}
\langle  ...\rangle =\frac{1}{A}\int_A d^2{\bf r} \,\, ...,
\end{equation}
where $A$ is the area to be sampled. In our case it will be area of the cavity,
 but in principle it
could be any area that one may wish to diagnose. In what follows we assume
that
 wave function $\psi({\bf r})$ is normalized as
\begin{equation}
\label{norma}
\int_A d^2{\bf r}|\psi({\bf r})|^2 =1,
\end{equation}
and therefore $\sigma^2 A=1$.
To bring  $\psi$ to a "diagonal" form in which the real and imaginary parts are
independent Gaussian fields we
introduce the new functions $p(x,y)$ and  $q(x,y)$ by changing the phase as

\begin{equation}\label{phasetrans}
\psi(x,y)\rightarrow e^{i\alpha}\psi(x,y)=p(x,y)+iq(x,y).
\end{equation}
The condition $\langle  pq\rangle =0$ now allows us to determine $\alpha$.
By this step we will also be able to
 find the analytic expressions for the
wave function and probability current statistics.
Straightforward algebra gives
$$
\tan2\alpha=\frac{2\gamma}{\sigma_u^2-\sigma_v^2},
$$
\begin{eqnarray}
\label{dispersions}
\langle  p^2\rangle =\frac{1}{2}\left[\sigma^2+\sqrt{\sigma^4-4(\sigma_u^
2 \sigma_v^2-
\gamma^2)}\right],\nonumber\\
\langle  q^2\rangle =\frac{1}{2}\left[\sigma^2-\sqrt{\sigma^4-4(\sigma_u^
2 \sigma_v^2-
\gamma^2)}\right].
\end{eqnarray}

\noindent Next let us consider
the cumulative distribution $G(\rho)$ for the scaled density
$\rho({\bf r})=A|\psi({\bf r})|^2$:
\begin{equation}
\label{cumul}
G(\rho)=\int_{C(\rho)}f(p,q)dp dq.
\end{equation}
The integration is defined by the circle $C(\rho)$ in the  $(p,q)$-plane
centered at the origin and with radius $\sqrt{\rho({\bf r})}$,
{\it i.e.}, $(p^2+q^2)/\sigma^2\leq \rho({\bf r})$
 in the integral above.
The function
 $f(p,q)$ is the joint distribution for the random Gaussian fields $p$ and $q$:
\begin{equation}
\label{fpq}
f(p,q)=\frac{1}{2\pi\sqrt{\langle  p^2\rangle \langle  q^2\rangle }}
\,\,e^{-(\frac{p^2}{\langle  p^2\rangle }+\frac{q^2}{\langle  q^2\rangle })/2}.
\end{equation}
After integration of Eq. (\ref{cumul}) we obtain
\begin{equation}
\label{cumul1}
G(\rho)=\frac{1}{2\pi}\int_0^{2\pi}\frac{1-\exp[-\rho\mu(\mu+\nu\cos\theta)]}
{\mu+\nu\cos\theta}d\theta,
\end{equation}
where we introduced the following notations
\begin{equation}
\label{munu}
\mu=\frac{1}{2}\left(\frac{1}{\epsilon}+\epsilon\right),
~\nu=\frac{1}{2}\left(\frac{1}{\epsilon}-\epsilon\right),
~\epsilon=\sqrt{\frac{\langle q^2\rangle}{\langle p^2\rangle}}\,.
\end{equation}
Differentiating (\ref{cumul1}) with respect  to $\rho$ we find the
final expression for the density distribution
\begin{equation}
\label{probdistr}
P(\rho, \epsilon)=\mu\exp(-\mu^2\rho)I_0(\mu\nu \rho),
\end{equation}
where $I_0(z)$ is the modified Bessel function of zeroth order.

The distribution in Eq.~(\ref{probdistr}) coincides
with the results obtained from RMT for closed systems
\cite{zyczkowski,kanzieper} and is
therefore not new. For weakly open systems with point contacts \v{S}eba {\it et al}.
\cite{sebaPRB} have related
 the statistical properties of the scattering matrix elements
with the distribution $P(\rho)$ and have obtained the expression above.
Analogous derivation of ours is also found in Ref.~\cite{RPnini}.
Our way of deriving Eq.~(\ref{probdistr}), however, explicitly shows
how to identify the two independent random fields in a given wave function.
For the random wave function in Eq.
(\ref{phasetrans}) the limits $\langle q^2\rangle \rightarrow 0,
\epsilon\rightarrow 0 $
correspond to a closed system and as a consequence one recovers
the P-T distribution and GOE statistics.
On the other hand, the case
$\langle p^2\rangle \rightarrow \langle q^2\rangle ,\epsilon\rightarrow 1$
corresponds in this context to an open system through which there is a
current flow. Consequently one finds
the exponential Rayleigh distribution
that corresponds to the GUE statistics.
In the crossover region, the value of $\epsilon$ is obtained numerically
using Eqs.~(\ref{disp}), (\ref{dispersion}), (\ref{dispersions}), 
and (\ref{munu}), {\it i.e.,
$\epsilon$ is not merely a fitting parameter}.
We have
recently verified this type of crossover in wave function statistics for
 a Bunimovich stadium using numerical scattering methods\cite{HIshio}.

In view of all previous work on the generic form of wave function statistics
in chaotic systems it is surprising
that no attention has been paid to the corresponding current distributions
except Ref.~\cite{RPnini}
where they only relate the average of squared current to $\langle\rho\rangle$.
Since currents may be measured \cite{MATopinka,PSeba} it is of interest
to establish
a form also for currents.
Below we will show how to find a useful form that is both simple
and universal. Let us limit ourselves to the case of a weak {\it net} current
between narrow input and output  leads. Inside the cavity, however, there will
be a rich, whirling flow pattern, which is strongly influenced
 by the vortical motions
around the nodal points associated with the complex form of the wave function.
Hence the net current through the through the billiard turns out to be only a 
tiny fraction of the total internal flow, in particular for asymmetric 
arrangements
 of leads and wave lengths that are small compared to the dimensions of the 
 cavity. As a result the corresponding distributions may
to a good approximation be chosen to be isotropic. Hence the components of the current effectively average to zero. These assumptions 
are verified by
the numerical calculations to be discussed in the next section.

Our complex wave function (\ref{psi})
carries  the probability current density ($\hbar=m=1$)
\begin{equation}
\label{current}
{\bf j}=\mbox{Im}(\psi^{*} \nabla \psi)= p\nabla q -q\nabla p.
\end{equation}

To find the corresponding
distribution it is convenient to begin with a
characteristic function for the components of the
probability current density
\begin{equation}
\label{char}
\Theta({\bf a})=\langle e^{i{\bf a\cdot j}}\rangle =
\langle \exp[i( p{\bf a \cdot \nabla} q-
q{\bf a \cdot \nabla} p)]\rangle .
\end{equation}
Since $\langle p\nabla q\rangle = \langle q\nabla p\rangle =0$ for isotropic 
fields $\nabla p$ and $\nabla q$ are statistically independent of $p$ and $q$.
They have the same distribution as in Eq. (\ref{fpq}) with dispersions
$\langle(\nabla p)^2\rangle = k^2 \langle p^2\rangle$ and
$\langle(\nabla q)^2\rangle = k^2 \langle q^2\rangle$ which follows from the
Schr\"odinger equation.
Using the relation $\langle ({\bf a}\nabla p)^2 \rangle = a^2 k^2 \langle p^2 \rangle/2$
and similarly for $\nabla q$ we obtain
\begin{equation}\label{char1}
\Theta(a)=\frac{1}{1+\tau^2\,a^2}\,,
\end{equation}
where $a=|{\bf a}|$ and
\begin{equation}\label{r}
\tau^2=k^2\langle p^2\rangle\langle q^2\rangle /2.
\end{equation}
From Eq. (\ref{char1}) it is
now easy to calculate the distribution functions. For
the components we have
\begin{equation}\label{jx}
\begin{array}{c}
P(j_x)=\langle\delta(j_x-p\frac{\partial q}{\partial x}
+q\frac{\partial p}{\partial x})\rangle=\\ [0.4cm]
\displaystyle\frac{1}{2\pi}\int_{-\infty}^\infty
\Theta(|a_x|)\,e^{-ia_x j_x}\,da_x=
\frac{1}{2\tau}\exp(-|j_x|/\tau)
\end{array}
\end{equation}
and the same for $P(j_y)$. In order to derive the
distribution function for the absolute value of the
probability current density let us consider the joint
distribution function
\begin{equation}\label{jxjy}
P(j_x,j_y)=\frac{1}{2\pi}\int_0^\infty
a\,J_0(aj)\Theta(a)\,da=
\frac{1}{2\pi\tau^2}K_0\left(\frac{j}{\tau}\right),
\end{equation}
where $j=|{\bf j}|$ and $K_0(z)$ is the modified
Bessel function of the second kind. Since this
expression is radially symmetric one can find the probability
density function $P(j)$ for $j$
 by just multiplying Eq. (\ref{jxjy}) with a factor of
$2\pi j$. This gives us the final expression
\begin{equation}
\label{absj}
P(j)=\frac{j}{\tau^2}K_0(j/\tau).
\end{equation}


\section{Numerical results}
As a numerical verification of the analytic expressions for the probability
current distribution,
we consider an open 2D Bunimovich hard-wall stadium (see inset in Fig.\ref{T}).
It is characterized by
the radius of a semicircle $a$ and the half-length of a straight section $l$,
and coupled to a pair of leads with a common width $d$.
Here we choose ($a=l$) and ($d/\sqrt{A}=0.0935$) for which the billiard
is maximally chaotic and weakly open, respectively.
To find the scattering wave function  for  particles entering and leaving
the cavity via the leads we solve the time-independent Schr\"odinger equation
for $\psi$
under Dirichlet boundary conditions
using a plane-wave-expansion method \cite{KNakamura}, which
gives reflection and transmission amplitudes for a given energy.
The wave functions are used to compute the different parameters
 entering the statistics using the explicit expressions stated above.

Figure \ref{T} shows the transmission probability $T$
as a function of Fermi wave number $k$ for an incoming wave
with transverse mode $n$ in the leads.
There is a sequence of overlapping resonances
which become broader in the high energy region shown in the
lower section of Fig.~\ref{T}.

For the statistical analysis of the scattering wave functions
we select two typical cases: (A) A low energy with only one fully open channel
($n=1$) for which $T$ reaches unity;
(B) A high energy with $n=4$ and an intermediate value for $T$. For the
statistics
the spatial average is taken over the billiard region
corresponding to the closed stadium. For convenience this area is set equal
to unity.

\vspace*{6.7cm}
\begin{figure}
\includegraphics{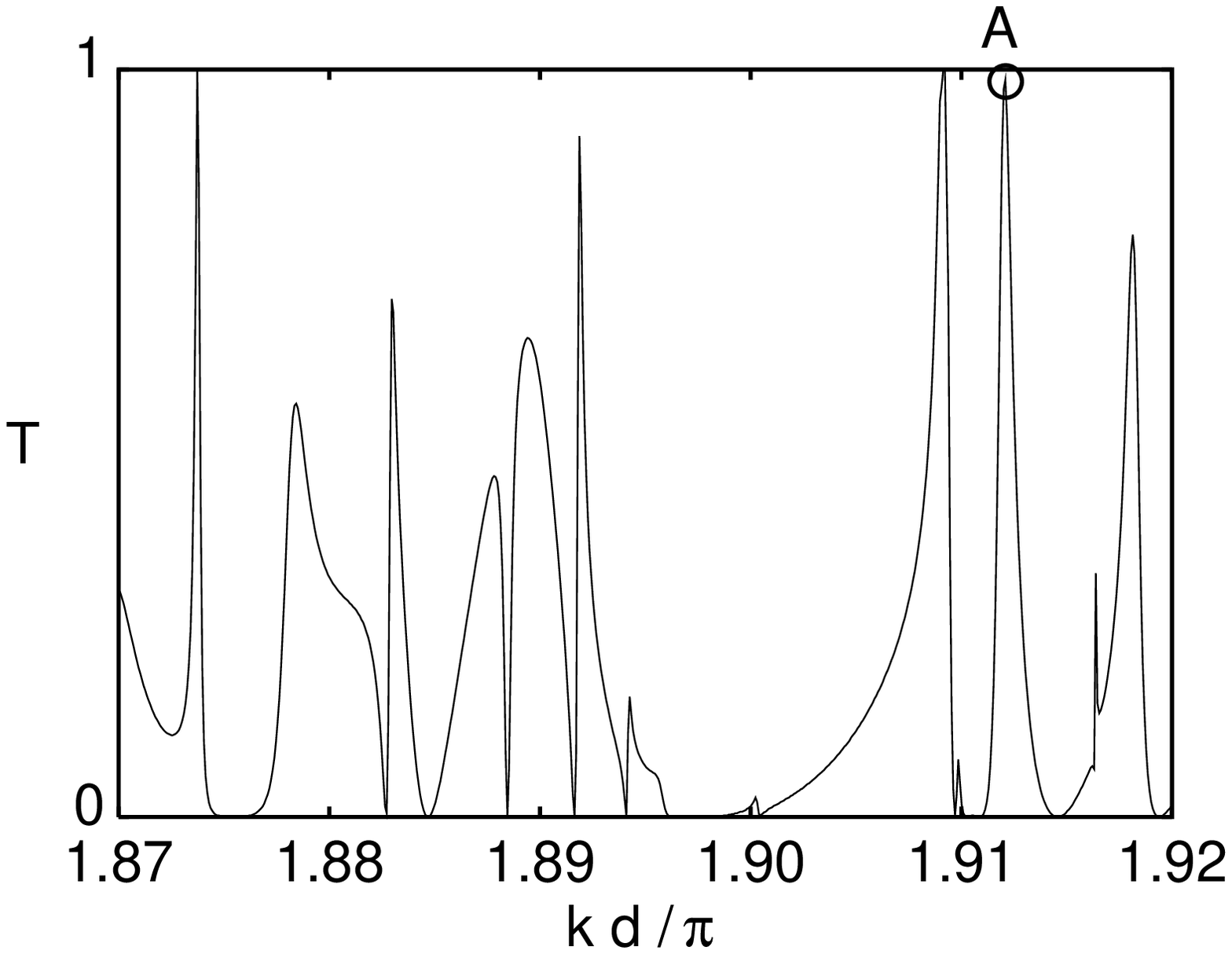}
\end{figure}
\vspace*{4cm}
\begin{figure}
\includegraphics{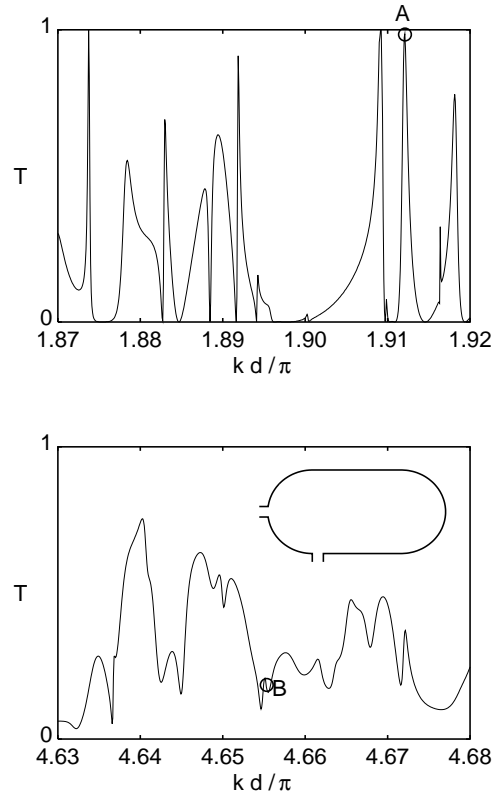}
\caption{Transmission probability $T$ as a function of Fermi wave number $k$
for the open stadium billiard:
(A) A low energy case with one open channel in the leads ($n=1$),
(B) A high energy case with $n=4$. The inset shows the hard-wall Bunimovich stadium and
the positions of the leads.}
\label{T}
\end{figure}

\begin{figure}
\includegraphics{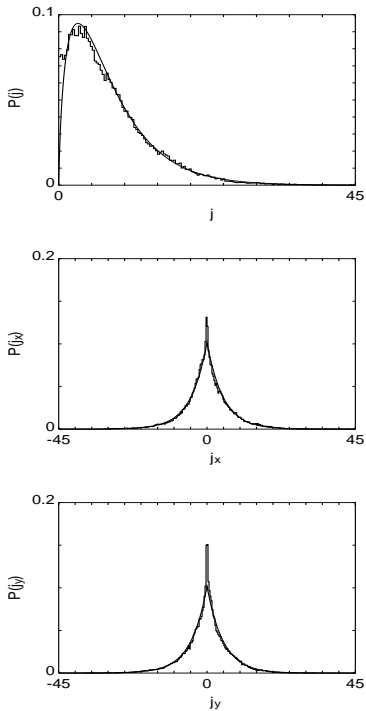}
\caption{Distribution of probability current density $P(j)$ (top)
and its components $P(j_x)$ and $P(j_y)$ (middle and bottom)
in the open stadium billiard for the case A
in Fig.~\ref{T}.
Solid curves show the analytical predictions
for $\epsilon=0.32$. (For convenience $\hbar=1, m=1$.)}
\label{Fig2}
\end{figure}

Figure \ref{Fig2} shows the numerical results
for $P(j)$, $P(j_{x})$ and  $P(j_{y})$ together with the analytical predictions
in Eqs. (\ref{jx}) and (\ref{absj}).
In the case A there is almost no reflection
and hence the system is completely coupled to the open channel.
The current statistics shows, however, that $\epsilon=0.32$, {\it i.e.},
intermediate between a closed and fully open cases.
Nevertheless, the numerical results show good agreement with the theory.

\begin{figure}
\includegraphics{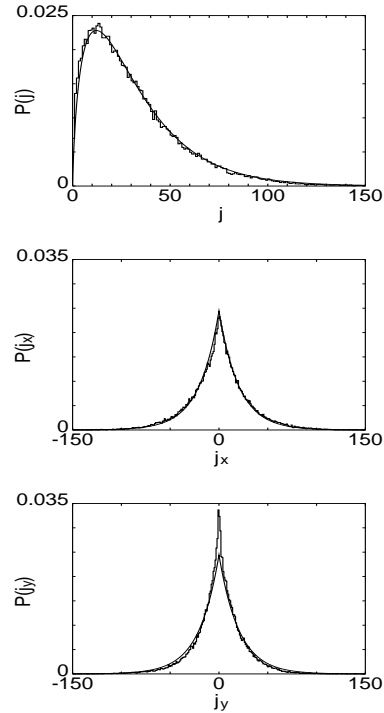}
\caption{Same as in Fig.~\ref{Fig2} but for the case B for which
$\epsilon=0.86$.}
\label{Fig3}
\end{figure}

Also in the high energy region B in Fig.~\ref{T}  the
probability current distributions are well described by the theory as shown in
Fig.~\ref{Fig3}.
Here $\epsilon=0.86$ which is close to the exponential Rayleigh (GUE) case
$\epsilon =1$.
The transmission is, however, lower than in the previous example.
As it appears from these
there is no simple relation between $T$ and $\epsilon$.
\vspace*{10.5cm}

\section{concluding remarks}

We have derived the statistical distributions for wave functions,
probability current densities and corresponding components
for 2D open quantum systems with classically chaotic dynamics.
The expressions for the probability currents are universal in the sense that
the shape of the distributions is independent of the mixing
parameter $\epsilon$, {\it i.e.}, only the width changes with $\epsilon$.
This is in contrast to the wave function statistics that transforms gradually
from GOE to GUE type with increasing $\epsilon$.
Obviously these ideas carry over into 3D.

The statistics for mesoscopic transport through a chaotic open billiard
was also studied numerically with enough statistical resolution to compare
with the analytical predictions. The discussed results give a numerical
proof of the predictions for the probability current density developed
here. It also appears that experimental verifications are
possible.
For example, images of the coherent electron flow
through a quantum point contact have been observed in recent experiments
\cite{MATopinka}.
There is also the case of thin microwave resonators \cite{HJStockmann,PSeba,barth}
and applying the present theory to
the Poynting vector.

\vspace*{10.5cm}

We acknowledge support from the Swedish Institute,
the Swedish Board for Industrial and Technological Development (NUTEK)
and the Royal Swedish Academy of Sciences. This work was also supported by
Russian Foundation for Basic Research (RFBR Grant 01-02-16077). Helpful 
discussions with M. Berry,
H.-J. St\"ockmann, and M. Barth are gratefully acknowledged (K.-F. B.).


\end{document}